\documentstyle[preprint,aps]{revtex}

\begin{document}
\tighten
\draft
\preprint{THU-94/09}
\title
{Nonlinear viscous vortex motion in
two-dimensional \\ Josephson junction arrays}

\author{T. J. Hagenaars$^1$, P. H. E. Tiesinga$^1$, J. E. van Himbergen$^1$
and Jorge V. Jos\'{e}$^{1,2}$ \\
{\it $^1$Instituut voor Theoretische Fysica,\\ Princetonplein 5,
Postbus 80006, 3508 TA Utrecht, The Netherlands\\
$^2$Department of Physics, Northeastern University,\\ Boston
Massachusetts 02115, USA}}
\date{\today}
\maketitle
\begin{abstract}
When a vortex  in a two-dimensional Josephson junction array is driven by
a constant external current it may move as a particle in a viscous
medium. Here we study the nature of this viscous motion.
We model the junctions in a square array as
resistively and capacitively shunted Josephson junctions and carry out
numerical calculations of the current-voltage characteristics. We find that
the current-voltage characteristics in the damped regime are well described
by a model with a {\bf nonlinear} viscous force of the form
$F_D=\eta(\dot y)\dot y={{A}\over {1+B\dot y}}\dot y$, where
$\dot y$ is the vortex velocity, $\eta(\dot y)$ is the velocity dependent
viscosity and $A$ and $B$ are constants for a fixed value of the
Stewart-McCumber parameter. This result is found to apply also
for triangular lattices in the overdamped regime.
Further qualitative understanding of the nature of the nonlinear friction on
the vortex motion is obtained from
a graphic analysis of the microscopic vortex dynamics in the array.
The consequences of having
this type of nonlinear friction law are discussed and compared to previous
theoretical and experimental studies.
\end{abstract}
\narrowtext
\maketitle

\newpage

\section{Introduction}
Understanding the vortex motion produced by
externally applied currents has been an important topic
in the study of the transport properties of Type II superconductors in the
Abrikosov phase. When vortices are able to move across the
system they produce Faraday voltages that are responsible for the I-V
characteristics measured experimentally. It has been useful to describe
the transport properties of the dilute vortex phase  in terms of a
phenomenological single vortex equation of motion. An
example of this approach is given by the Bardeen-Stephens equation which
has been successfully applied to conventional superconductors \cite{tinkham}.
A similar approach has been attempted in the description of the transport
properties in two-dimensional Josephson junction arrays (JJA)
\cite{Korea,Los,EckSch,Eck,Rzch,OrlM,Dick,EckSon,Anne,Herre1,Tighe,HHerre,Bobb,UliG,YuS,YuS2}.
These arrays are 2D lattices of superconducting islands (sites)
connected by Josephson junctions (bonds).
The unit cells (plaquettes) of these lattices can be, for
example, square or triangular.
In the JJA case the vortices are represented by eddy current patterns
about a plaquette.
Although these JJA vortices differ in several important
ways from their continuum counterparts, the question that has
been addressed by several authors is: To what extent can one use
a single macroscopic equation of motion to describe the
dynamical properties  of vortices
in JJA? Further interest in this problem has come from
recent experiments in  underdamped arrays \cite{Herre1,Tighe,HHerre}.
These arrays were found to show hysteretic features in their I-V
characteristics that suggest that vortices behave as
particles with a mass \cite{Herre1}.
Furthermore, experimental evidence for ballistic vortex motion was
reported in triangular arrays \cite{HHerre}.

In this paper we concentrate on the friction experienced by a
JJA vortex. We investigate this friction in detail by
numerical simulation of the dynamics of an array
containing one single vortex. The commonly adopted vortex
equation of motion assumes a frictional force proportional to the vortex
velocity. Our results show that, instead, the friction is a
nonlinear function of the vortex velocity that decreases as the velocity
increases. We propose a new phenomenological friction law that accounts for
the numerical results.

Here we consider the classical regime defined by $E_{J}>>E_{c}=e^2/2C$,
where $E_J$ is the Josephson coupling energy and $E_c$ the
charging energy of two islands, $e$ the electron charge,
and $C$ the capacitance of the  junction.
In this regime quantum fluctuations are neglected, leaving the phases
$\theta (\bbox{r})$ of the Ginzburg-Landau order parameter on the
islands as the only dynamical variables.
The experiments mentioned above were reported to be in this regime.
In this case the JJA are
well-modelled by the RCSJ model, defined by the total bond
current $i(\bbox{r},\bbox{r'})$, between nearest  neighbor
sites $\bbox{r}$ and $\bbox{r'}$
\begin{eqnarray}
i(\bbox{r},\bbox{r'})&=&
\beta_{c}\ddot{\theta}(\bbox{r},\bbox{r'})\nonumber \\
&+& \dot{\theta}(\bbox{r},\bbox{r'})
+\sin[\theta(\bbox{r},\bbox{r'})
-2\pi A(\bbox{r},\bbox{r'})],\label{RSJ}
\end{eqnarray}
plus Kirchoff's current conservation conditions at each site.
Here the dots represent time derivatives. The three contributions
to  $i(\bbox{r}, \bbox{r'})$ are the  displacement, the dissipative and
the superconducting currents, respectively.
The phase difference across a junction is
$\theta(\bbox{r},\bbox{r'})\equiv
\theta(\bbox{r})-\theta(\bbox{r'})$.
The currents are expressed in
units of the junction critical current $I_{c}$;
time is measured in units of the characteristic time
$1/\omega_{c}=\hbar/(2eR_{n}I_{c})$, and
$\beta_{c}=(\omega_{c}/\omega_{p})^2$ is the Stewart-McCumber
parameter \cite{StMc}, with the plasma frequency $\omega_{p}$ defined as
$\omega_{p}^2=2eI_{c}/\hbar C$,  and $R_n$
is the junction's normal state resistance.  The bond frustration variable
$A(\bbox{r},\bbox{r'})$ is defined as the line integral of the vector
potential $\bbox{A}$:
\begin{equation}
A(\bbox{r},\bbox{r'})=\frac{1}{\phi_{0}}
\int_{\bbox{r}}^{\bbox{r'}}\bbox{A}\cdot d\bbox{l},
\end{equation}
with the elementary quantum of flux $\phi_{0}=hc/2e$.

It has been suggested that the transport properties emerging from Eq.
(\ref{RSJ}),
in the case that the phase configurations in the array contain one vortex,
can be described by a classical macroscopic model in terms of a single
continuous vortex
coordinate $y$ that satisfies the equation of
motion \cite{EckSch,Eck,Rzch,OrlM,EckSon}
\begin{equation}
M\ddot{y}+\eta \dot{y}=i_{b}+i_{d}\sin(2\pi y),\label{eom1}
\end{equation}
where $M=\pi\beta_{c}$ and $\eta=\pi$ for
a square array, and
$M=2\pi\beta_{c}$ and $\eta=2\pi$ for a triangular array.
This equation assumes that the JJA vortex can be described as a point
particle with mass $M$ that, driven by a (Lorentz) force $i_{b}$,
moves through a sinusoidal
pinning potential and experiences a viscous damping force with constant
viscosity coefficient $\eta$. The vortex  mass $M$ can be
calculated \cite{Rzch} by equating
the electromagnetic energy stored
in the array to a vortex kinetic
energy $\frac{1}{2}M\dot{y}^2$.
The value of the depinning current, $i_d$,
depends  on the underlying lattice geometry. An estimate for $i_d$ in
a square lattice gives $i_d\approx 0.1$  while for a triangular lattice
it is only $i_d \approx  0.02$ \cite{LAT}.

If one substitutes $2\pi y$ by $\theta$ in Eq. (\ref{eom1}), one obtains
the equation of motion for the phase difference across
a single Josephson junction
with current bias $i_{b}$, critical current $i_d$,
shunt resistance $2R_{n}$ (square
lattice) or $R_{n}$ (triangular
lattice) and shunt capacitance $C/2$ (square lattice) or $C$ (triangular
lattice).
Eq. (\ref{eom1}) has been studied extensively, mostly numerically, and the
results show
different types of nontrivial behavior
depending on the values of the parameters in the equation \cite{StMc}.
The solutions to Eq. (\ref{eom1}) exhibit a critical value $i_b=i_d$
above which the junction is in a nonzero voltage state. In the
array case $i_d$ corresponds to
the depinning current above which the vortex moves and a finite
voltage is measured. This voltage is proportional to the vortex velocity
and arises because the phase differences across the array
change in time.
When $M=0$, the overdamped case, and for
currents $i_b<i_d$ the vortex is pinned to the lattice and the voltage
is zero. For $i_b>i_d$ the vortex can move under the action of the current
and a nonzero voltage state is produced.
For $M\neq 0$ the I-V characteristics resulting from Eq. (\ref{eom1}) show
hysteretic behavior when the current $i_b$ is ramped up and down past
the depinning current $i_d$.
If $M$ is sufficiently large Eq. (\ref{eom1}) predicts ballistic
vortex motion, in the sense that a high-velocity vortex would
continue its motion over many lattice constants when the
driving current is switched off.
Van der Zant {\em et al.} \cite{HHerre} reported experimental
evidence of ballistic vortex motion
in a region without driving currents
in a H-shaped triangular array with
$\beta_c=46$.
The quoted $\beta_c$ value was computed from the normal state
resistance of the junctions. At low temperatures and voltages, the
effective $\beta_c$, determined by the quasi-particle resistance, can be
orders of magnitude larger \cite{HHerre}.
In contrast to this experimental result,
in numerical simulations within the RCSJ model,
no evidence for ballistic vortex motion has been
found\cite{Bobb,UliG,YuS,YuS2};  a high velocity vortex in an underdamped array
does not move more than one plaquette as soon as the driving current
is switched off. Furthermore, the calculated I-V characteristics
for square arrays show almost no hysteresis near the vortex depinning current
for $\beta_c=10$ \cite{YuS}, whereas Eq. (\ref{eom1}) would yield a substantial
hysteretic behavior.

In trying to understand this discrepancy between experimental
and theoretical studies an additional dissipative mechanism, arising from
the coupling of the vortex to spin waves or plasma oscillations,
has been shown  to give rise to a nonzero vortex viscosity in the completely
underdamped limit ($\beta_{c}=\infty$) \cite{UliG}. This would invalidate
the model (\ref{eom1}) in this limit, leading to very small
mean free paths over which  vortices come to rest if the driving current
is switched off,  even in highly underdamped arrays.
The enhanced viscosity has also been measured experimentally in
Ref. \cite{Herre1}. Nevertheless is was suggested there that the
vortices might still move ballistically in a current-free region
at low velocities.

In this paper we carry out a systematic comparison between the results for
the I-V characteristics obtained from simulations of JJA described by Eq.
(\ref{RSJ}) and the
I-V characteristics obtained from an equation of the form given in Eq.
(\ref{eom1}). The analysis is
carried out for a range of $\beta_{c}$ values.
We find that an equation of the form of Eq. (\ref{eom1}) is not
representative of the JJA results. Instead we find strong quantitative
evidence that an equation that yields a rather good fit to the
JJA results is
\begin{equation}
M(\beta_c)\ddot{y}+\frac {A(\beta_c)}{1+B(\beta_c)\dot{y}}\, \, \, {\dot{y}}
=i_{b}+i_{d}\sin(2\pi y).\label{non}
\end{equation}
Here the constants $A$, $B$ and $M$ are found to be weakly dependent
functions of $\beta_c$. This is the main result of this paper. We note that
the linear friction law given in Eq. (\ref{eom1}) has to be modified in a
nonlinear
way to account for the JJA results.  This nonlinear dependence on the vortex
velocity applies in particular to the range
$\beta_c=0$ up to $\beta_c\approx 100$.
This change to a nonlinear dissipation law raises some important questions,
for example, how to introduce temperature effects at the
phenomenological level \cite{Nico}. We will discuss other important
consequences
emerging from this nonlinear viscosity law later in the paper.

The outline of the paper is as follows. In Section II we discuss the
calculational algorithm used to compute the I-V characteristics
from Eq. (\ref{RSJ}).
In Section III we present the bulk of our results for the I-V characteristics
together with the fitting analyses that lead to the result given in
Eq. (\ref{non}).  In this section we also discuss
the microscopic aspects of the vortex motion in the array by analyzing the
current distributions of the vortex as a function of time.
Section IV contains our conclusions together with
a comparison to previous experimental and theoretical work.

\section{\bf {CALCULATIONAL APPROACH}}

In this paper we are interested in calculating the dynamical response of
an array of Josephson junctions driven by a constant d.c. current.
The set of nonlinear
dynamical equations of motion given in Eq. (\ref{RSJ}) can be efficiently
integrated using a fast Fourier transform algorithm \cite{DickFFT}.

In our simulations we use a square lattice (with a lattice
constant set equal to unity) with periodic boundary conditions
(pbc) along the $y$-direction while the current is fed in and taken out
along the $x$-direction (see Fig. 1). Hence a vortex tends to
move in the $y$-direction.
The total number of plaquettes along
the $x$- and $y$-directions
are denoted by  $N_{x}$ and $N_{y}$, respectively, whereas the total number of
sites is $L_{x}\times L_{y}$, with $L_{x}=N_{x}+1$ and $L_{y}=N_{y}$.
Most of the results presented in this paper correspond to systems
with $L_{x}=L_{y}$. However, these
results do not change significantly when considering systems
with $L_{x}\leq L_{y}$. In fact, if $L_{x}$ is not smaller than 8,
the weak finite size effects encountered in our calculations
are mainly governed by the vortex motion along the ${y}$-direction.

The vorticity $n(\bbox{R})$ of a plaquette $\bbox{R}$ can be defined as
(see for instance the Appendix of Ref. \cite{UliG})
\begin{equation}
2\pi n(\bbox{R})=2\pi f +\sum_{{\cal P}(\bbox{R})}
\big(\theta(\bbox{r},\bbox{r'})-2\pi A(\bbox{r,r'})\big).
\label{vortdef}
\end{equation}
Here ${\cal P}(\bbox{R})$ denotes an anticlockwise sum around
the plaquette $\bbox{R}$ and the gauge invariant phase difference
$\theta(\bbox{r},\bbox{r'})-2\pi A(\bbox{r},\bbox{r'})$ is
taken between $-\pi$ and $+\pi$.
The frustration parameter $f$ measures the average flux piercing a
plaquette, measured in units of $\phi_{0}$.
Physically, vortices in JJA can be seen as eddy currents in the
current flow pattern.
If there is only one vortex in the array, then there is one plaquette,
say $\bbox{R}_{0}$, with
vorticity  $n(\bbox{R}_{0})=1$ while all other plaquettes have
zero vorticity. We will call $\bbox{R}_{0}$ the core of the vortex.
In a JJA with pbc the phase configurations corresponding to
a single vortex in the middle column of the array cannot be written down
as easily as in an array with free boundaries. We construct a single vortex
configuration with a method used previously in Ref. \cite{Dick}.
It allows for a direct calculation of the phase configuration in
terms of the vorticities $n(\bbox{R})\in \{-1,0,1\}$ once a
gauge choice for the $A(\bbox{r},\bbox{r'})$ and a choice for one of
the phases $\theta(\bbox{r})$ has been made.
Here we are mainly interested in understanding the one-vortex
dynamics and thus we concentrate on this case throughout the paper.

We take the frustration equal to $f=1/N_{x}N_{y}$ so that the
single vortex we introduce in the middle column of the array has
a current pattern symmetric around that column.
The single vortex equation of motion proposed in Eq. (\ref{eom1})
or Eq. (\ref{non}) describes a
continuous  motion in the $y$-direction,
whereas the location of a vortex as determined from the phase configurations
is discrete  and  undetermined within  the vortex core.
Therefore, when making comparisons between the vortex
velocity, defined in terms of the microscopic phases, to that obtained from
the coarse grained vortex variable $y$, we need to compare time averaged
quantities.  The vortex velocity is directly related to the time average
of the voltage $V(t)$ across the array in the $x$-direction,
where $V(t)$ is defined as
\begin{equation}
V(t)=\sum_{y=0}^{L_y-1}\frac{d}{dt}[\theta (L_x-1,y)-\theta (0,y)],
\end{equation}
according to the Josephson relation.
Time is again measured in units of $1/\omega_{c}$,
and $V(t)$ is measured in units
of $R_{n}I_{c}$.
Each time the vortex has
travelled over a distance $N_{y}$ the total phase
difference across the array has changed by $2\pi$ and therefore
the vortex velocity $\it v$ is given by
\begin{equation}
v= \frac{1}{2\pi}V,\label{vovo}
\end{equation}
where $V=<V(t)>$.

Another quantity of interest in describing the vortex dynamics is the
current vorticity around a plaquette defined as
\begin{equation}
C(\bbox{R},t)=\sum_{{\cal P}(\bbox{R})}i(\bbox{r,r',t}).
\end{equation}
An important difference between the vorticity $n(\bbox{R},t)$ and
$C(\bbox{R},t)$ is that the former is an integer, while the latter is a
continuous function describing the vortex as an eddy current pattern
extending outside the vortex core.
One interesting quantity to look at is the
``center of mass" of the current vorticities
\begin{equation}
\tilde{Y}_v\equiv
\frac{\sum_{\bbox{R}}R_{y}C(\bbox{R})}{\cal N}=
\frac{\sum_{R_{y}}R_{y}D(R_{y})}{\cal N}
\label{comdef}
\end{equation}
with
\begin{equation}
D(R_{y})=\sum_{R_{x}}C(\bbox{R}).
\end{equation}
The normalization factor ${\cal N}$ has an unusual form depending on the
pbc assumed in our calculations. It is determined
by the requirement that $\tilde{Y}_v$ has to change one unit if
the current vorticity configuration is shifted one plaquette.
For two current vorticity configurations $C( \bbox{R})$
and $C'( \bbox{R})$ shifted by
exactly one plaquette with respect to each other, we get
\begin{eqnarray}
{\cal N}&\equiv&
\sum_{R_{y}}R_{y}D'(R_{y})-\sum_{R_{y}}R_{y}D(R_{y})\nonumber
\\
&=&\sum_{R_{y}}D(R_{y})-L_{y}D(L_{y}-1).
\end{eqnarray}
By taking large enough lattice sizes, one can ensure that
the quantity $\cal N$ is essentially constant
for a range of positions of the vortex core in the middle of the coordinate
system.
In this region $\tilde{Y}_{v}$ show steps with integer height magnitudes.
\section{\bf {RESULTS}}
In this section we present the evidence we have found that allows us to
conclude that a vortex in a JJA moves with a nonlinear viscosity law,
at least in the overdamped ($\beta_c =0$) to damped
($\beta_c\leq 100$) regime.
We start by considering the overdamped case, in which there are
no shunt capacitors,  and therefore there is no spin-wave dissipation
channel.  Next we will discuss the results for
values of $\beta_c$ up to 100.

\subsection{Nonlinear viscosity in the $\beta_c=0$ case}

Fig. 2 shows a typical I-V characteristic computed for
a 32$\times$32 array with one vortex ($f=1/N_{x}N_{y}=1/992$) and with
$\beta_{c}=0$. In order to compare with the
result from Eq. (\ref{eom1}), we have plotted the
vortex velocity as defined in Eq. (\ref{vovo}) versus the current.
The results for the I-V characteristics
for larger lattices are practically the same, as we will
discuss in more detail below.
For currents $i_b\leq 0.10$, the vortex is pinned
by the lattice and thus the measured voltage is zero. Above $i_b \approx 0.10$,
the
vortex is depinned from the lattice  by the current and its motion gives
rise to a nonzero voltage. Up to approximately $i_b=0.97$, the
time-averaged vortex velocity gradually increases and so does the
measured voltage across the array. This is the regime where the
phenomenological equation of motion must apply and thus
we call the range $i_b \in [0,0.97)$ the vortex regime.
At $i_b\geq 0.97$ we enter a current regime in which
eventually all individual junctions in the current direction perform
phase slips.

In Fig. 2 we also show, as a continuous line, the result for the I-V
characteristic assuming
the validity of Eq. (\ref{eom1}),  with the identification
of the parameters as deduced in Refs. \cite{Eck,Rzch,EckSon}. In the $M=0$ case
considered
here the known analytic expression for the I-V characteristic is given by,
\begin{equation}
v=\frac{1}{\int_{0}^{1}\frac{1}{\dot{y}}dy}=
\frac{\sqrt{i_{b}^2-i_{d}^2}}{\pi}.
\end{equation}
In Fig.2 we observe that the vortex equation of motion seriously
underestimates the time-averaged voltage almost everywhere in the
vortex regime. There is quantitative agreement with the calculated
results only for bias currents very close to the depinning current. More
interestingly, we observe a ${qualitative}$
difference between the two curves. Whereas the viscosity $\eta$ in Eq.
(\ref{eom1}) is constant  the simulations show an effective viscosity
which  ${decreases}$ with increasing vortex velocity.
This leads us to propose as a model a vortex equation of motion
of the form ($\beta_c=0)$,
\begin{equation}
\label{phe1}
\eta(\dot{y}) \dot{y}=i_{b}+i_{d}\sin(2\pi y),
\end{equation}
with a velocity-dependent viscosity $\eta=\eta (\dot{y})$. We have found
that the functional form for $\eta(\dot y)$ that fits our results
in the vortex regime quite well reads,
\begin{equation}
\eta(\dot{y})=A/(1+B\dot{y}).\label{phe2}
\end{equation}
where the sign of $\dot{y}$ is taken positive and $A$ and $B$ are parameters
determined by fitting the array results to
this form. As in the constant viscosity case we can analytically evaluate
the result for the I-V characteristics yielding,
\begin{equation}
v= \frac{\sqrt{i_{b}^2-i_{d}^2}}{A-B\sqrt{i_{b}^2-i_{d}^2}}.
\label{IVform}
\end{equation}
The top curve in Fig. 3  shows a fit
using Eq. (\ref{IVform})
in the current range $i_b=0.10$ to $i_b=0.80$,
to the $\beta =0$ I-V characteristic
obtained from the simulations
of a $32\times 32$ array.
The values for the parameters are $A=2.67$
and $B=1.80$.
To indicate the error bars of these values, we mention that,
if we fit the form (\ref{IVform}) to the simulation results in
the range $i_{b}=0.10$ to $i_{b}=0.70$,
the fitted values for $A$ and $B$ are approximately
$0.5\%$ and $2.0\%$ larger, respectively.

To check on a possible size dependence of these results we carried
out the same analysis for lattices of sizes
$64\times 64$ and $128\times 128$. We find that the I-V
characteristics leads to
essentially the same results for currents in the range
$i_{b} \in (0.10,0.80)$. Representative results from the
fits for $16\times 16$ and $32\times 32$ arrays
are given in the inset of Fig. 3. The results for the two larger sizes
are essentially indistinguishable, within their error
bars, from the ones shown for $32\times 32$ arrays.

We conclude at this stage that the $\beta_c=0$ results are rather well
described
by the phenomenological Eqs. (\ref{phe1}) and (\ref{phe2}). In the inset of
Fig. 5 we show
the comparison between the friction force $F_D$ in the constant viscosity model
(Eq. (\ref{eom1})) and the one
proposed here (Eq. (\ref{non})) using the $\beta_c =0$ values for the
parameters $A$ and $B$ derived from
our fits.

\subsection{Nonlinear viscosity in the $\beta_c\neq 0$ case}

We move now to discuss the changes that occur when $\beta_c\neq 0$.
We solve the phenomenological vortex equation of motion, Eq. (\ref{non}),
numerically
to compare to the results obtained from solving the JJA equations.
In Fig. 3 we show the results for the I-V characteristics of a
$32\times 32$ array in the vortex regime
for $\beta_c=0, 1.2, 2, 3$ and $25$. We also show the fits to the
array results using the form given in Eq. (\ref{IVform}).
Nonlinear behaviour persists for values
of $\beta_c$ up to 50. The fits for small $\beta_c$ are
of the same quality as for $\beta_c =0$.
Note that we take $M(\beta_c)=0$ in these
fits, although Eq. (\ref{eom1}) suggests a non-zero mass $M$ as soon as
$\beta_c\neq 0$.
Including a mass $M$ as a parameter, as in equation (4), does
not
result in a better fit of the nonlinearity in the vortex
regime.
The choice $M(\beta_c)=0$ is corroborated by the fact that
we do not find any measurable hysteresis
near the depinning current in the simulated I-V characteristics
for Stewart-McCumber parameters even up to $\beta_c\approx 35$,
in agreement with the results for $\beta_c=10$
reported previously in Ref. \cite{YuS}.
The puzzling conclusion is then that, even when
the microscopic equations of motion have a ``mass" term, the
phenomenological vortex equation of motion behaves as if the vortex mass is
zero or very small. In the inset of Fig. 3 we show the $\beta_c$ dependence
of the parameters $A$ and $B$ for $\beta_c$ up to 4.
We note that $A$ increases slightly with $\beta_c$ while $B$ decreases
slightly . This trend indicates that the viscosity becomes ``more" linear
as $\beta_c$ increases.
This trend can be understood as being a consequence of the ``spin-wave
friction" mechanism that sets in at $\beta_c >0$ and leads to an enhancement
of the linear viscosity. An approximate estimate
of the  linear viscosity in this regime presented in Ref. \cite{UliG}
led to a rise  roughly proportional to $\sqrt{\beta_c}$.
The same trend was found experimentally and a semi-quantitative
explanation of the results was given in  the second reference of \cite{Herre1}.

For lattices of size $8\times 8$ and larger
one needs $\beta_c$ values of the order of 100 to detect
small hysteresis loops near the depinning current.
For these lattice sizes, no hysteresis is
measured up to $\beta_c=35$ in our , using a current grid
as small as
$3\times10^{-6}$.
We have found, however, that for a small array of size $4\times4$,
a very small hysteresis loop is visible in this $\beta_c$ regime
which  resembles in shape
the ones obtained using Eq. (\ref{eom1}).
In Fig. 4 we show these hysteresis loops for $\beta_c$ values
between 7 and 13. Note that all the I-V characteristics have the same depinning
current
while ramping the current up whereas they have different zero voltage
intercepts when lowering the currents.

In Fig. 5 we show the vortex regimes of the I-V characteristics
for $\beta_c$ values up to 100 on a $32\times 32$ lattice.
Here we only show the result from ramping up the current,
thereby omitting the small hysteresis loops below the depinning
currents for $\beta_c=50$ and 100.
For $\beta_c\geq 7$, we find sharp jumps in the voltage.
The vortex regime ends at these jumps, which are believed to be due
to switching of rows of junctions to the resistive state.
This row-switching behavior has been seen before in experiments
\cite{Herre1,Tighe,HerreR} and in simulations \cite{Bobb,UliG,YuS,YuS2,YuR}.
In this figure we note a crossover from a nonlinear to a linear viscosity
regime as $\beta_c$ increases.
At $\beta_{c}=100$ the vortex regime of the I-V characteristic is
nearly linear for $i_b>0.25$ and can be extrapolated through the origin
(the step-like structure of the I-V characteristics in
the upper half of the vortex regime corresponds to
interference of the vortex with its periodic image, as was
explained in Ref. \cite{UliG}, and will disappear if
we consider a system with larger $L_{y}$).
For $\beta_c=50$ a similar extrapolation does not
intersect the origin, so the friction is  still nonlinear. We note that
the range of applicability of the nonlinear viscosity model given in Eq.
(\ref{non})
covers  some of the $\beta_c$ values reported in the experiments
in Refs. \cite{Herre1}.

\subsection{Nonlinear viscosity in triangular arrays}

All of the calculations described above were performed in square lattices.
Recently Yu and Stroud carried out calculations of the I-V characteristics in
triangular arrays \cite{YuS2}. We have reanalyzed their results in light
of our nonlinear viscosity model given in Eq. (\ref{non}).
In Fig. 6 we show the corresponding fit
to their $\beta_c=0$ results for the I-V
characteristic
of an $8\times 8$ lattice with the current biased in the $[10\bar{1}]$
direction.  The fits to
the parameters $A$ and $B$ yield the results
$7.67$ and $2.47$, respectively. From these numbers we conclude
that the magnitude of the viscosity is roughly
2.8 times as large as in the square lattice case,
while the nonlinearity parameter $B/A$ is smaller.
We note that a theoretical prediction of a factor of two for $A$
between the square and triangular lattices was made in Ref.
(9).

\subsection{Microscopic vortex motion}
It is interesting to directly study the time evolution of the vortex motion
across
the array in order to further understand the nature of the nonlinear
viscosity suggested by the phenomenological macroscopic model given in Eq.
(\ref{non}). We concentrate here on the $\beta_c=0$ case.

In Fig. 7(a) we show current vorticity distributions of a vortex
for $i_b=0.11$, slightly above the depinning current, at different times.
We observe that the vortex motion has essentially two time scales,
a slow and a fast one. In the slow regime
the vortex does not move
much while it gets deformed by the applied current.
Subsequently
the vortex moves fast until it gets stuck  again and the stretching
process repeats itself. This type of stick-slip-like motion is
reflected in the nonlinearity of
the viscosity. The decrease of the viscosity with increasing
velocity is analogous to the behaviour of the kinetic friction
coefficient between dry surfaces in the stick-slip phase, which is
likely to be generic for frictional
dynamics at low speeds \cite{baum}.
As shown in Fig. 7(b)
the qualitative motion of the vortex remains the same
for $i_b =0.6$, although the quantitative values for the slow and fast times
have become
smaller.

In Fig. 8 we show results for the time-dependent voltage $V(t)$ (full line)
and  the normalized center of mass vortex velocity (dashed line)
defined as
\begin{equation}
\tilde{\it {v}}(t)=
\frac{d}{dt}\tilde{Y}_{v},\label{oormerk}
\end{equation}
where $\tilde{Y}_{v}$ is given in Eq. (\ref{comdef}), for
three values of $i_b$. In order to ensure a constant value for the
normalization
factor  $\cal N$ in a wide range of vortex positions we choose a lattice
of size $8\times 64$.
We observe that in both the $V(t)$- and the $\tilde{\it v}(t)$-curves,
the amplitude of the oscillations around the average value decreases
with increasing the bias currents.
If we interpret the quantity $\tilde{\it v}(t)$ as a
coarse-grained vortex velocity, the physical meaning of this
result is that the pinning force decreases when the vortex
velocity increases.
To check this interpretation we extract the pinning barrier from the
simulations, by measuring the variation in the array energy given
by
\begin{equation}
E=\sum_{<\bbox{r,r'}>}\cos[\theta({\bbox{r,r'}})
-2\pi A({\bbox{r,r'}})]\label{ener}.
\end{equation}
We find that the pinning barrier shows a similar decrease as a
function of the (time-averaged) vortex velocity.
Under the assumption that the pinning force in the array is
proportional to the pinning energy barrier, we conclude that indeed
the amplitude of the pinning force decreases when the
vortex velocity increases, in accordance with the
interpretation of the quantity $\tilde{\it v}$.
We can qualitatively relate this result to the current vorticity
snapshots shown in Fig. 7:
for larger currents the vortex moves faster, the
current vorticity spreads out over more plaquettes and the pinning at the
core plaquette becomes less effective.

\section{\bf {CONCLUSIONS AND COMPARISON TO PREVIOUS WORK}}

In this paper we have proposed a phenomenological vortex equation of
motion that fits well the I-V characteristics obtained from
solving the full set of JJA microscopic dynamical equations. The main
difference with previous studies is that our proposed equation of motion
has a nonlinear velocity dependent viscosity that decreases as the
velocity increases. The validity of this description covers the range
from overdamped to damped regimes as defined by the Stewart-McCumber
parameter, and it applies to square as well as to triangular lattices.
We also have found that for $\beta_c \leq 35$ the
the I-V characteristics indicate that the vortex moves as if its inertial
mass is zero, or at most very small (no hysteresis at depinning).
As $\beta_c$ increases the nonlinearity of the viscosity
slowly decreases at the same time that the linear term slowly increases.
We will now discuss the above results in the light of previous experimental and
theoretical studies.

Experimentally, evidence for the nonlinear viscosity can be seen in
the I-V characteristics reported in Refs. \cite{Herre1,Tighe}.
However, the I-V characteristics results measured in
Ref. \cite{Herre1} for an almost overdamped triangular array
do not show evidence for a nonlinear viscosity, whereas in our simulations
it is in the overdamped case that the nonlinearity is dominant (see
Fig. 5).

On the theoretical side, Eckern and Sonin \cite{EckSon} derived a
general vortex equation of motion in the continuum limit.
In the adiabatic, or small vortex velocity limit,
this equation reduces to the model of
Eq. (\ref{eom1}) without the sinusoidal pinning force.
This equation of motion is believed to take into account
the spin-wave friction occurring when $\beta_c\neq 0$, as
found in Ref. \cite{UliG}.
Here we will focus on the $\beta_c=0$ case, in which
the equation of motion also provides  corrections to Eq.
(\ref{eom1}) beyond the adiabatic limit.
Taking a constant vortex velocity $v$, for
a constant current bias, for $\beta_c=0$
the vortex equation of motion reduces to
\begin{equation}
\label{eomA}v\int d^2 k\frac{k_{x}^2}{k^2}
\frac{e^{-k/\sqrt{2\pi}}}{1+v^2k_{y}^2}=2\pi i_{b}.
\end{equation}
Here the integral in $k$-space is over the two-dimensional
plane.
The vortex velocity  $v$ is taken along the $y$-direction.
The exponential in the integrand provides a smooth cutoff for
large $k$. An alternative cutoff used in Ref. \cite{UliG}
consists of replacing the exponential in the integrand
by the two-dimensional Heaviside function
\begin{equation}
\Theta(\;\mid\!k_x\!\mid-\pi)\Theta(\;\mid\!k_y\!\mid-\pi).\label{Heaviside}
\end{equation}
Note that in Eq. (\ref{eomA}) we cannot add a sinusoidal pinning
force, as this would be inconsistent with the constant-velocity assumption.
However, the inclusion of the pinning potential, being most important in
producing a finite
depinning current, would barely affect the higher-velocity part of the I-V
characteristics, where the nonlinearity in Eq.
(\ref{eomA}) is most pronounced.

In Fig. 9 we show the I-V characteristics
computed from Eq. (\ref{eomA}), with the exponential $k$-cutoff, and
for the cutoff given in (\ref{Heaviside}).
Both curves intersect the origin, because the lattice
pinning potential
is absent in (\ref{eomA}). We note that the inclusion of
non-adiabatic effects in this equation of motion
gives rise to a viscosity that decreases with increasing velocity.
Although there is an improvement in the higher-velocity part of
the I-V characteristic when compared to the linear viscosity vortex
equation of motion, Eq. (\ref{eom1}), the predictions of
the continuum model still deviate qualitatively from the
full (lattice) calculations. We note that the higher-velocity component
of the I-V characteristic depends crucially on the choice of high-momentum
cutoff in  Eq. (\ref{eomA}).

This work has been motivated in part by the issue of
ballistic vortex motion.
The phenomenological vortex equation of motion
presented in this paper
attributes a mass $M(\beta_c)=0$ to the vortex
in square arrays
in the regime of $\beta_{c}\leq 35$.
This is a consequence of the absence of hysteresis in the
I-V characteristic in this regime, also reported in Ref. \cite{YuS}.
This means that the electromagnetic energy stored in the
shunt capacitors does not represent a kinetic energy
for the vortex in this regime, at least not in
the way suggested by the
model Eq. (\ref{eom1}). This detracts from the idea behind the
possibility of ballistic vortex motion in JJA described by the
RCSJ model.

In a separate argument, the enhancement of the viscosity
with increasing $\beta_c$,
leads to very small path lengths
over which a vortex with high initial velocity
looses its assumed kinetic energy. The enhancement of the viscosity was
also measured experimentally \cite{Herre1}. It has been explained
in Refs. \cite{Herre1,UliG}
in terms of an additional
friction mechanism
due to coupling of the vortex to plasma oscillations.
It was also suggested in Ref. \cite{EckSon} that this coupling
would not prevent ballistic vortex motion in a small
velocity window in triangular arrays. Recent
simulations \cite{YuS2} of triangular arrays did not
show such a velocity window in the parameter range considered
($0\leq\beta_c\leq 1000$).
However, one may need much larger values of $\beta_c$ to
possibly see ballistic vortex motion \cite{HHerre}.

For the discrepancy between the
results of the experiment of Van der Zant et al. \cite{HHerre}
and that of
the simulations based on the RCSJ model,
one possible explanation
suggested recently in Ref. \cite{Anne}
involves the discreteness of the charges  in  the array.
The clarification of this problem needs further experimental and
theoretical study.
With regard to the nonlinear vortex viscosity found in
our work, establishing a direct connection between the microscopic
and the phenomenological description represents a difficult problem
for future study.

\acknowledgments
We thank  U. Geigenm\"{u}ller for useful suggestions about the paper
and we thank him and H. van der Zant and J. B. Sokoloff
for illuminating conversations.
This work was part of the research program of the ``Stichting
voor Fundamenteel Onderzoek der Materie (FOM)", which is financially
supported by the ``Nederlandse organisatie voor Wetenschappelijk
Onderzoek (NWO)". The work of JVJ was partially supported by
$NSF$ grant DMR-9211339.


\newpage

\begin{figure}
\caption{Array geometry used in the simulations, illustrated
with an $8\times 8$ array. Junctions are denoted as crossed bonds.
In the $y$-direction periodic boundaries are
imposed, while the current bias is applied  along the $x$-direction.}
\label{fig1}
\end{figure}

\begin{figure}
\caption{$\beta_c=0$ I-V characteristics, plotted as average
vortex velocity versus normalized bias current. The dashed line
gives the results from simulations of a $32\times 32$ array with one vortex.
The simulation shows a vortex viscosity that decreases with
increasing vortex velocity. The full line was obtained from the model
vortex equation of motion (3). }
\label{fig2}
\end{figure}

\begin{figure}
\caption{ Simulation results for the I-V characteristics as in Fig. 2
for different values of $\beta_c$ (circles).
The full lines are the fits to the I-V characteristics using Eq.
(15)
for a $32\times 32$ lattice.
For clarity of presentation the origin of successive
$\beta_c$ values is offset to the right by 0.1 unit.
The inset shows the values of the fitted parameters $A$ and $B$ as
a function of $\beta_c$ and array size. Diamonds ($A$) and triangles ($B$)
correspond to $32\times 32$ array whereas circles ($A$) and squares
($B$) to $16\times 16$.}
\label{fig3}
\end{figure}

\begin{figure}
\caption{Hysteresis loops in the simulated
I-V characteristics for a $4\times 4$  array
with one vortex, for different $\beta_c$ values.
Note the smallness of the  current scale. First the current is swept up
to $i_b=0.1265$, slightly above the
depinning current, and subsequently
it is swept down until the vortex is
retrapped by the lattice.
The depinning current is
$i_{b}=0.126$ for all
$\beta_{c}$ values shown. For
$\beta_c=7$ the current at which the vortex is retrapped
is (on this scale) equal to
the depinning current.  For higher values of $\beta_c$ this
current is increasingly lower. For $8\times 8$ and larger lattices the
hysteresis
for these $\beta_c$ values disappears.}
\label{fig4}
\end{figure}

\begin{figure}
\caption{I-V characteristics (vortex
velocity versus bias current) from
simulations of a $32\times 32$ array with one
vortex, for different values of $\beta_c$. From top
(marked with `1') to bottom (`2'),
$\beta_c=0,1.2,3,7,15,50,100$,
respectively. Note that the higher $\beta_c$, the smaller the vortex regime.
In the inset the nonlinear friction force
$F_D(\dot{y})=A\dot{y}/(1+B\dot{y})$ (full line) is shown as a
function of the vortex velocity $\dot{y}$, for the $\beta_c=0$
parameter values $A=2.67$ and $B=1.80$. The dashed line
is the friction force $F_{D}(\dot{y})=\pi\dot{y}$
in the constant-viscosity model (3).}
\label{fig5}
\end{figure}

\begin{figure}
\caption{Simulation results from Ref. [17] for an $8\times 8$ triangular array
with current along
the [10\={1}] direction, $\beta_c=0$ (circles).
The continuous line is the fit using Eq. (15).
The fit parameters obtained are
$A=7.67$ and $B= 2.47$.}
\label{fig6}
\end{figure}

\begin{figure}
\caption{Snapshots of a smooth interpolation of one vortex
current vorticity distribution  in an $8\times 8$ sublattice
of a $16\times 16$ array,
 for two different current values: (a) $i_b=0.11$, (b) $i_b=0.60$.
Different gray scales represent different levels
of current vorticity. In the first snapshots (labelled as
`0'), the black dot is the middle point of the vortex. The
dashed line (fixed in time) is a guide to the eye.
In (a) the time interval between two snapshots is
$\Delta t=10$ (in units of $1/\omega_c$).
The vortex moves over one plaquette in approximately
$t=7 \Delta t$.  In (b) the time interval between successive snapshots
is $\Delta t=0.625$ in units of $1/\omega_c$, and here the period of the
motion is approximately $t=4.5\Delta t$.}
\label{fig7}
\end{figure}

\begin{figure}
\caption{Rescaled voltage  (full lines) and vortex center of mass
velocity (dashed lines) versus time,
in an $8\times 64$ array, for three different bias currents.
The amplitude of the oscillatory
component in both quantities decreases with increasing $i_b$.}
\label{fig8}
\end{figure}

\begin{figure}
\caption{Comparison of the
$\beta_c=0$  average vortex velocity versus
normalized bias current obtained from
a simulation of a $32\times 32$ array with
one vortex (dashed line) with the one from Eq. (18)
(curve (a)).
Curve (b) is obtained from Eq. (18) by replacing
the smooth high-momentum cutoff in (18) by the sharp
cutoff (19). Curve (c) is obtained from the model vortex equation of
motion (3). }
\label{fig9}
\end{figure}

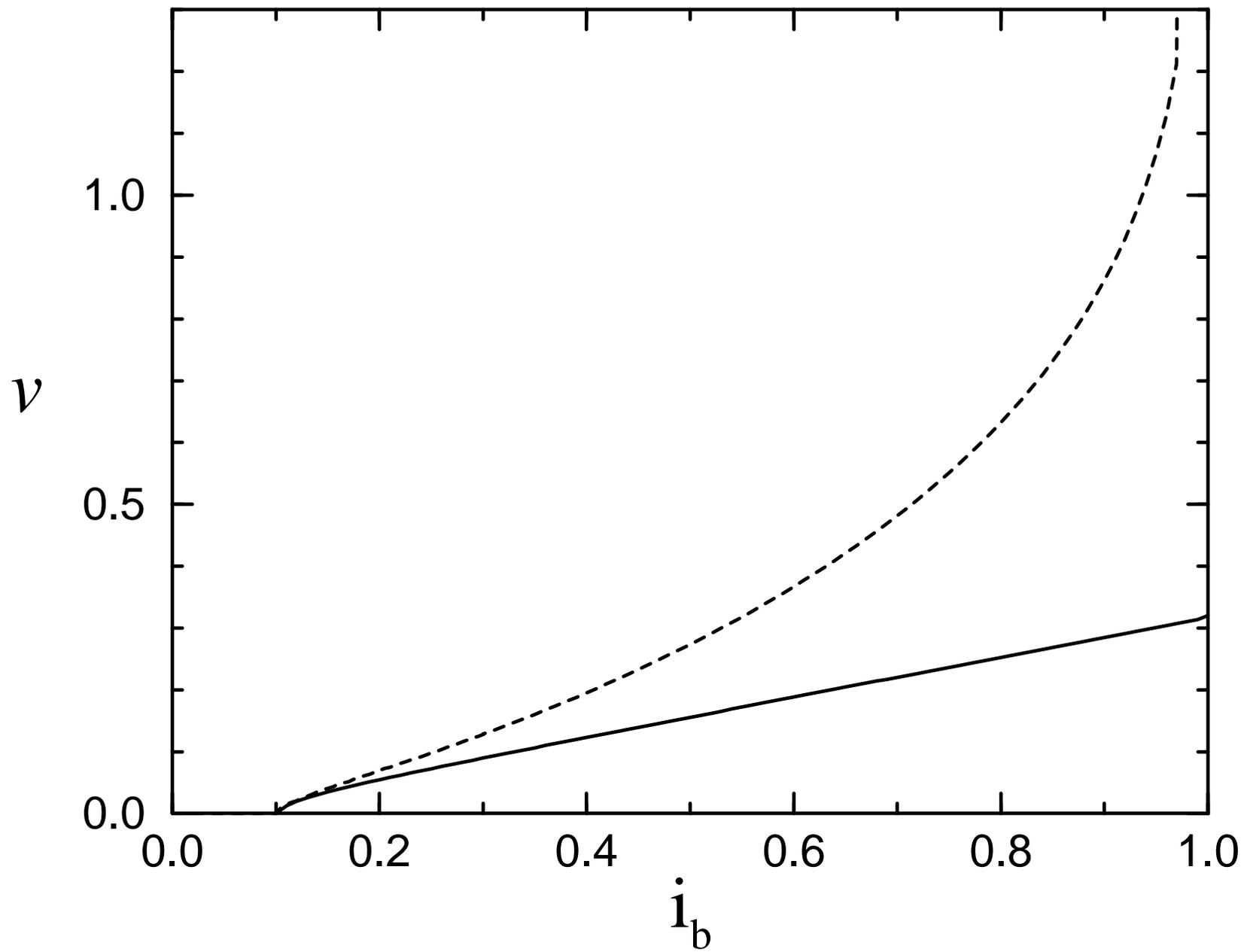
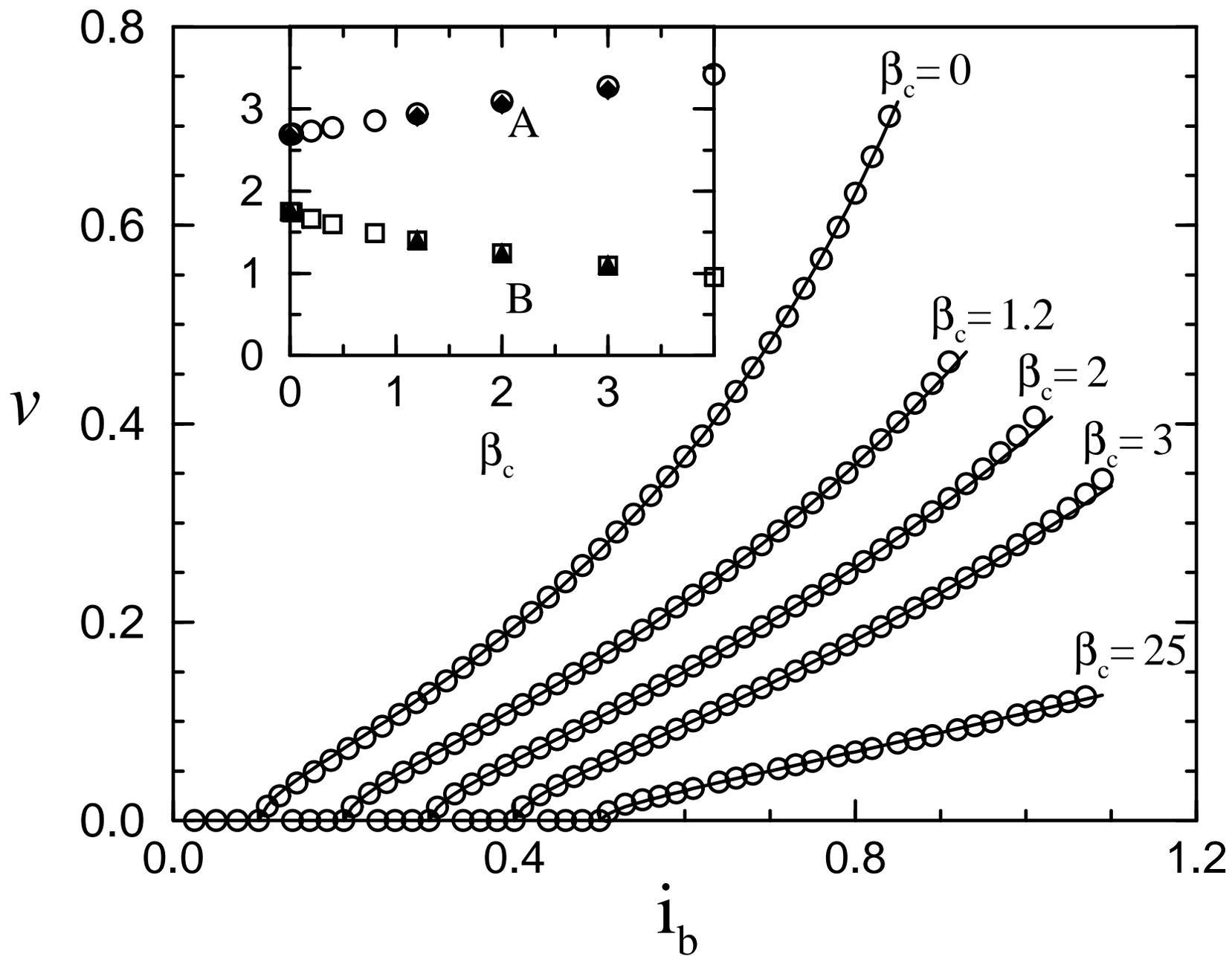
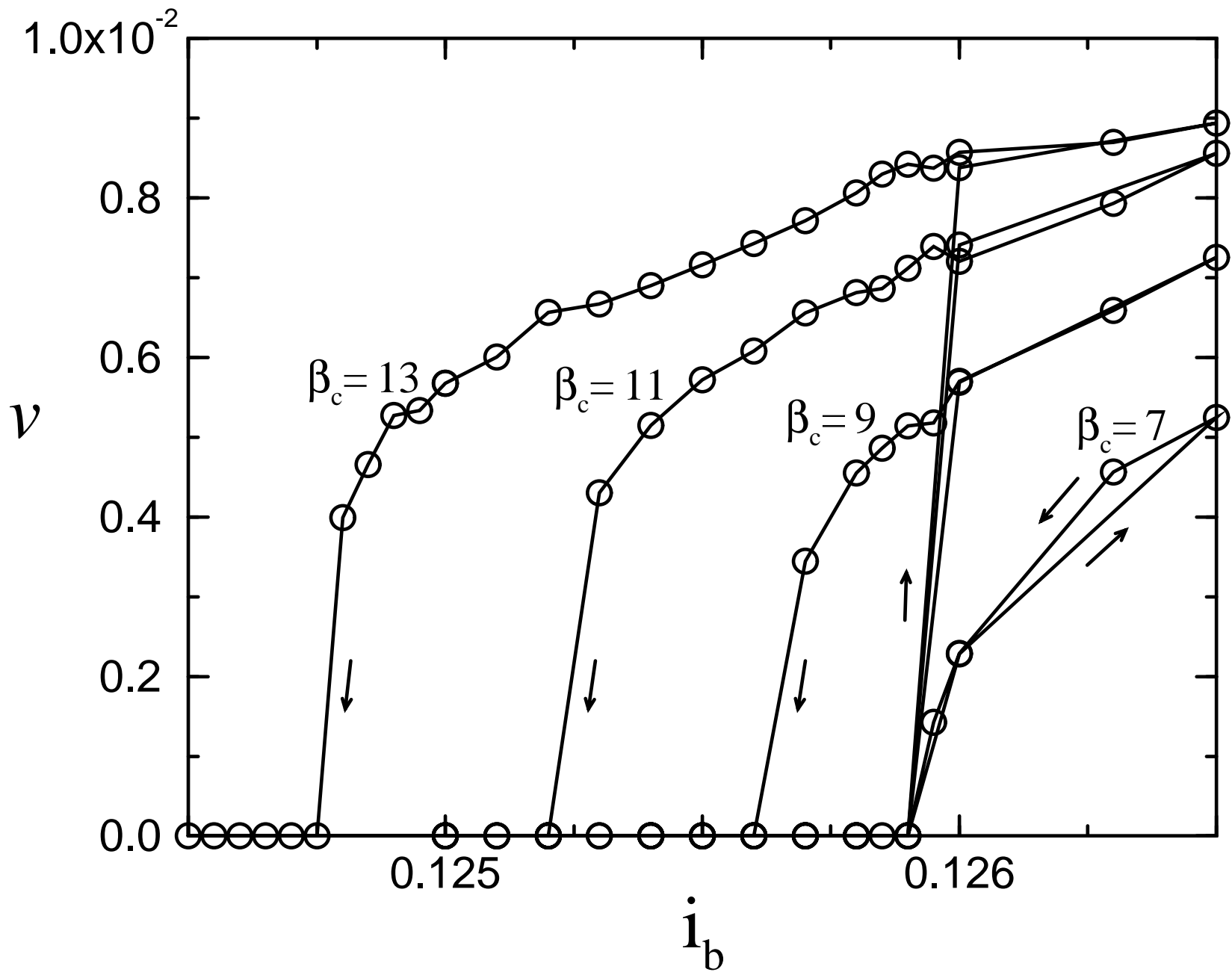
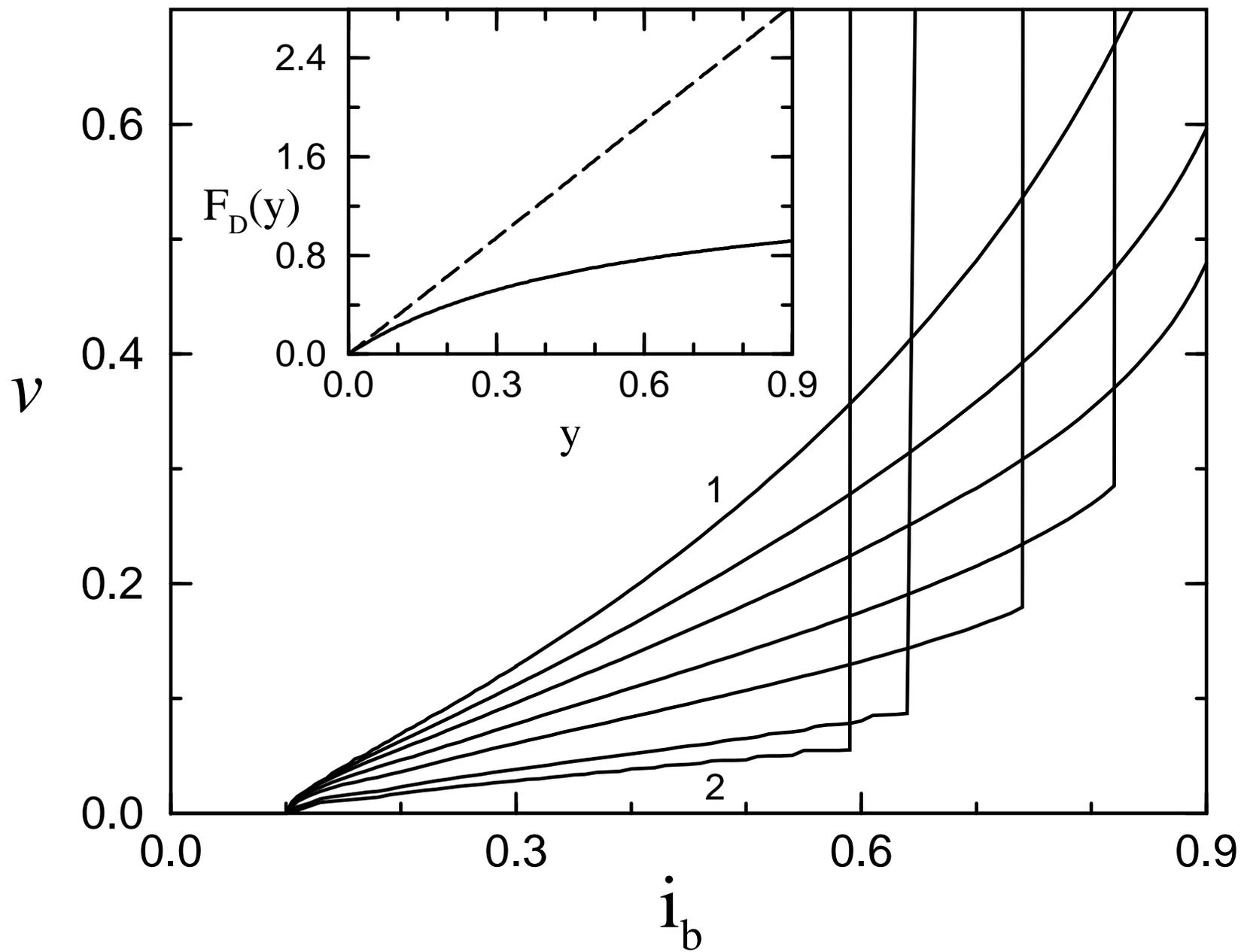
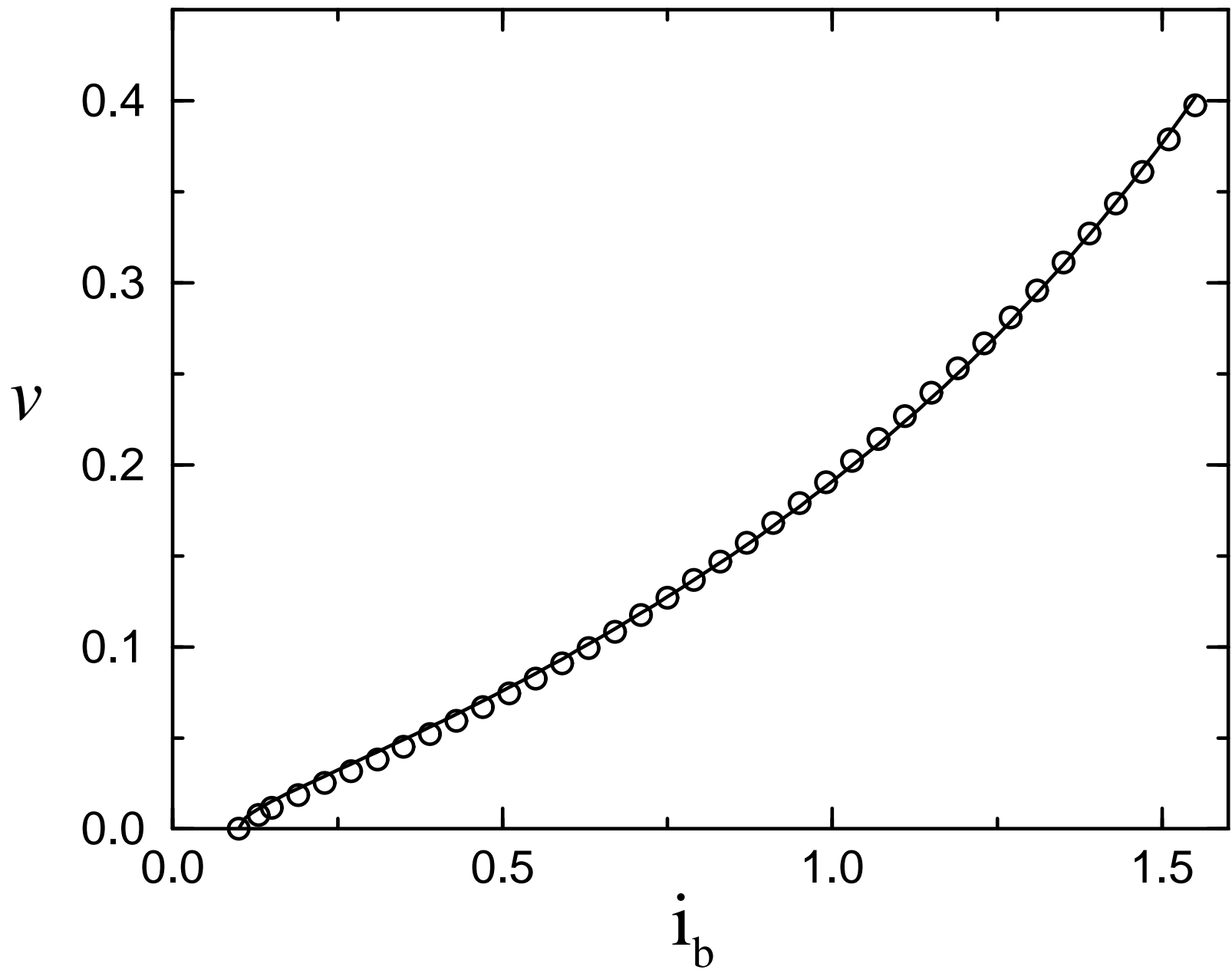
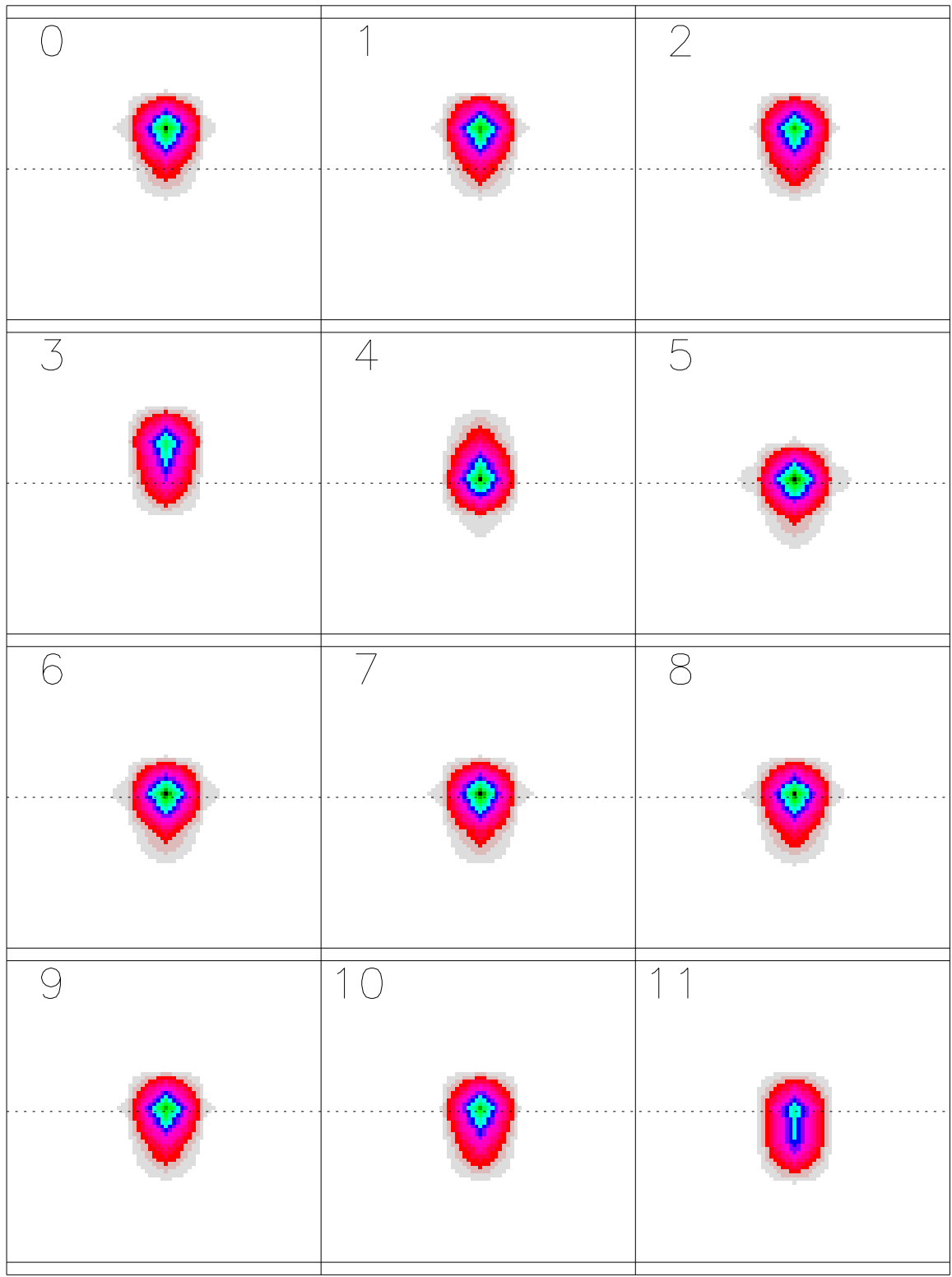
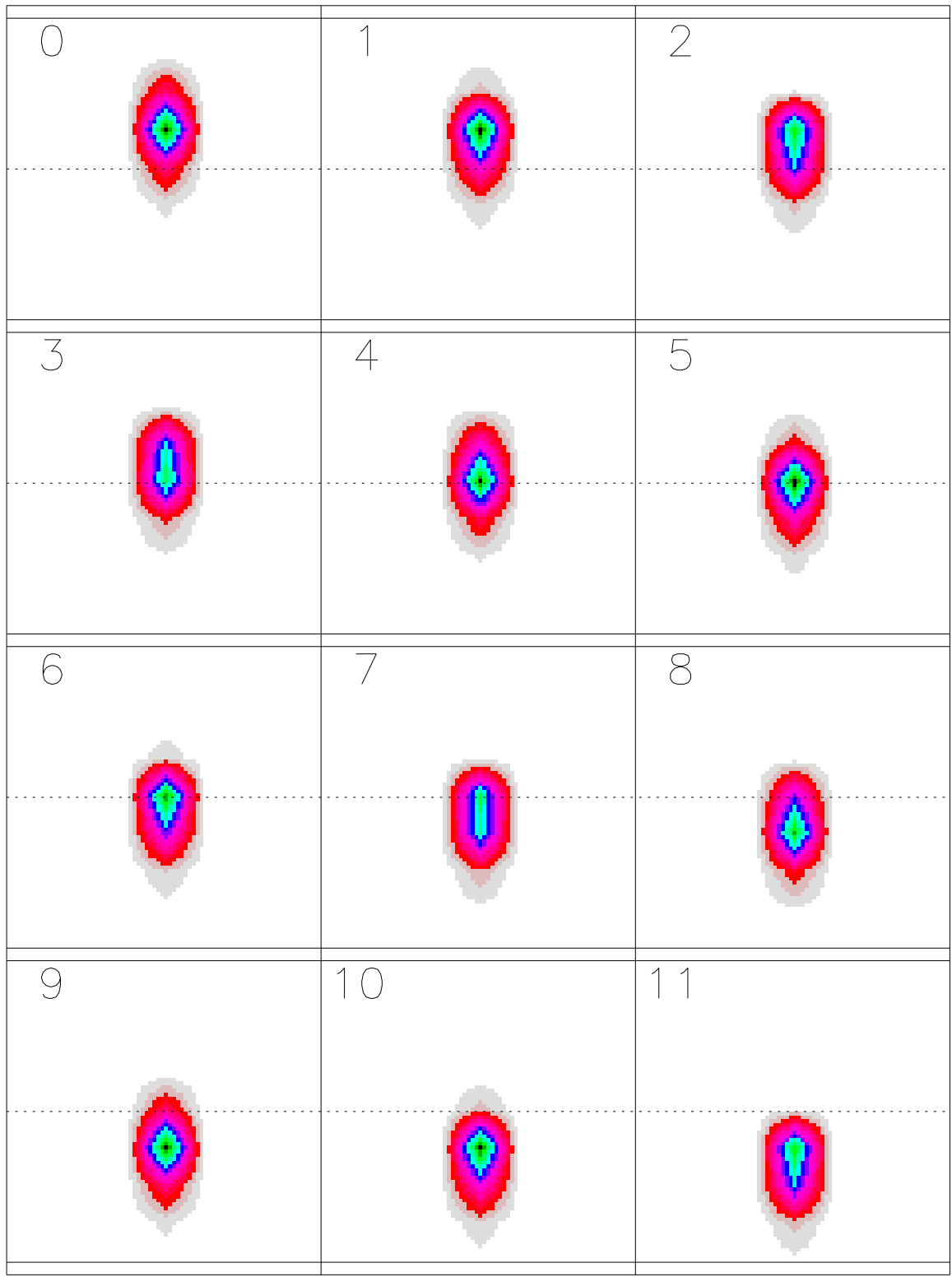
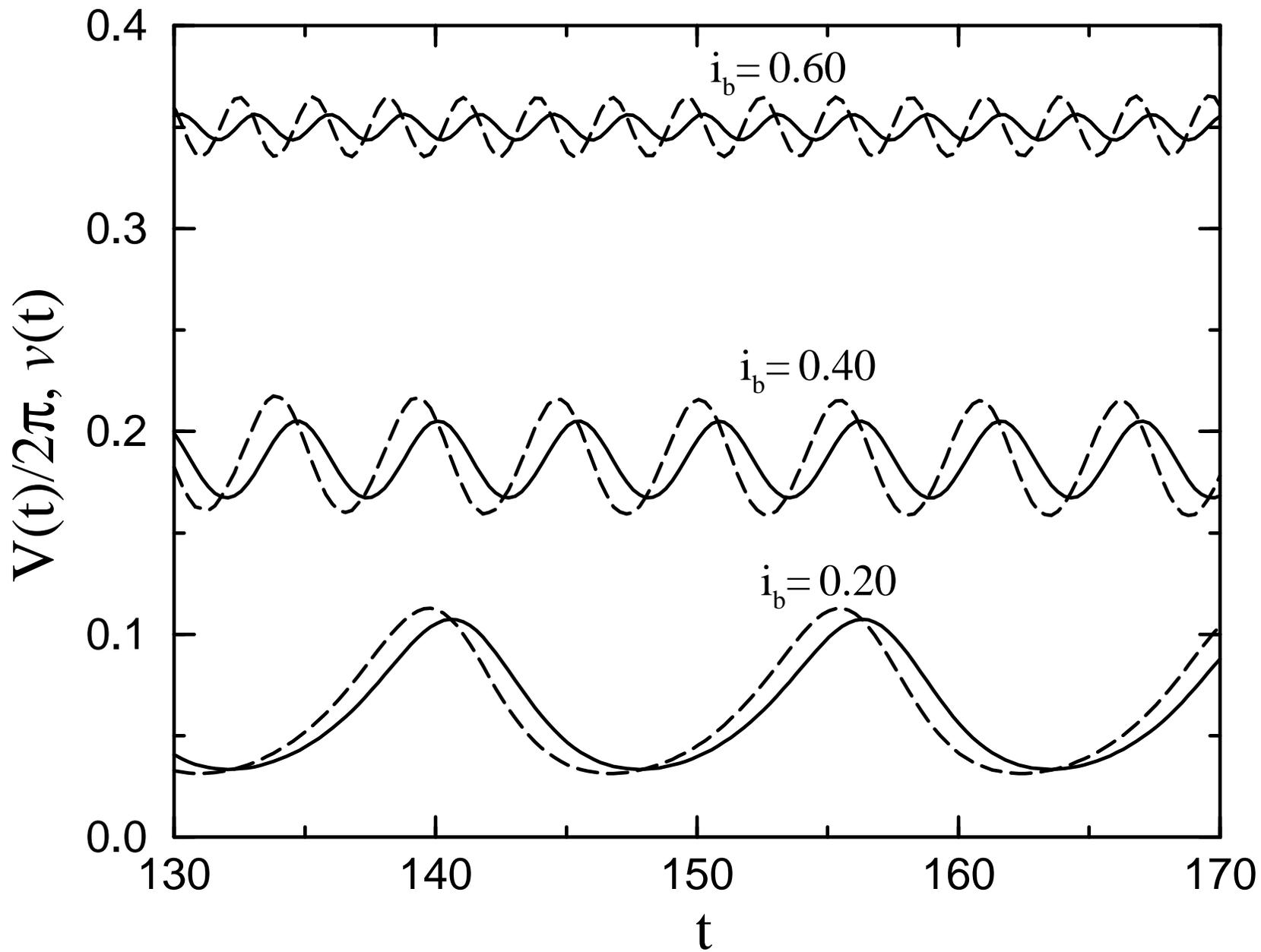
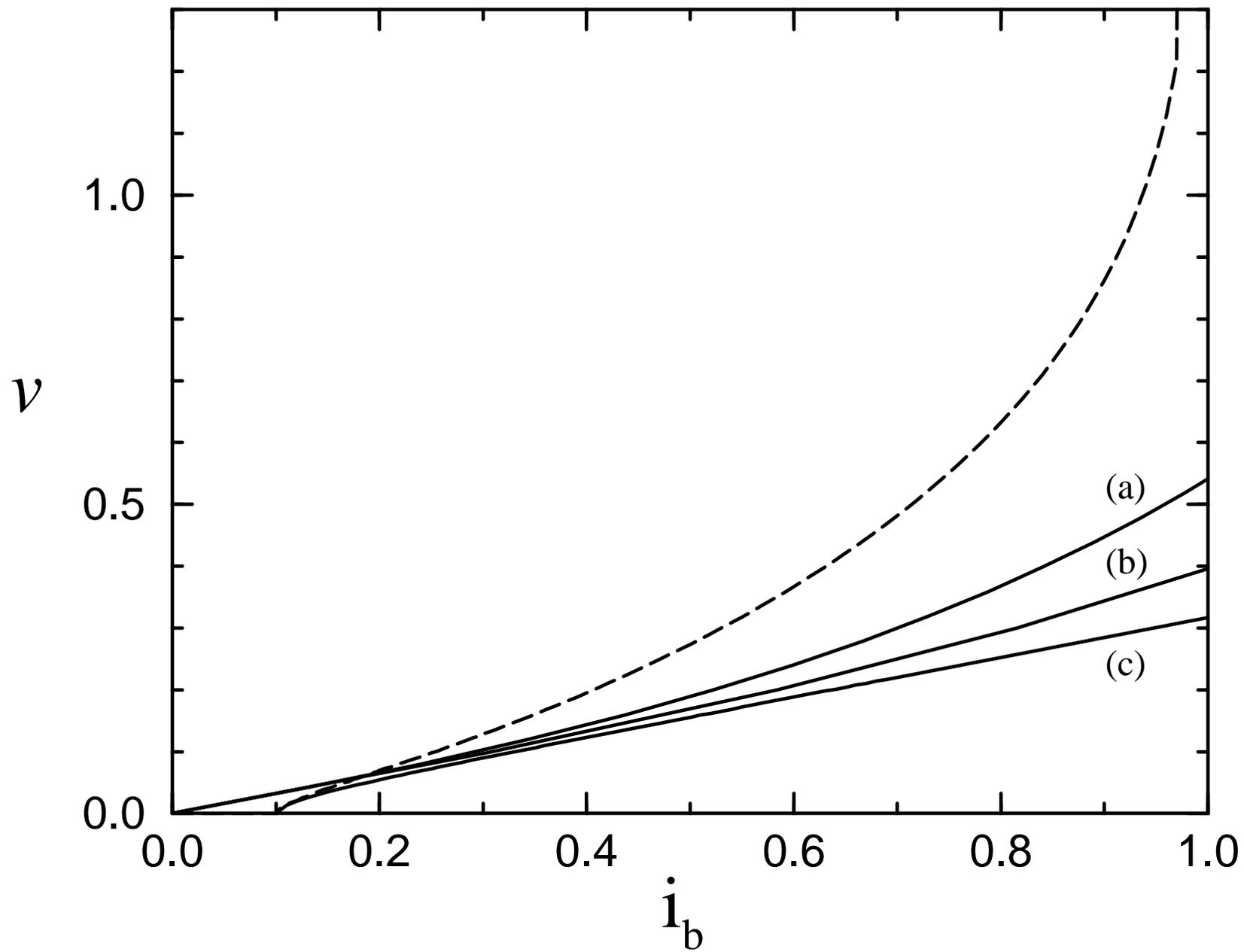
\end{document}